\documentclass[journal]{IEEEtran}
\usepackage{graphicx}
\usepackage[T1]{fontenc}
\usepackage{lmodern}
\usepackage{textcomp}
\usepackage{latexsym}
\usepackage{amssymb}
\usepackage{epsfig}
\usepackage{url}
\usepackage[fleqn]{amsmath}

\newcommand{\ee}        {{\rm e}}
\newcommand{\vct}[1]    {\mbox{\boldmath{$#1$}}}
\newcommand{\jj}        {{\rm j}}
\newcommand{\dd}        {{\rm d}}
\newcommand{\diag}        {{\rm diag}}

\newcommand{\HT}        {{\rm H}}
\newcommand{\TT}        {{\rm T}}
\newcommand{\EE}        {{\rm E}}
\newcommand{\cum}        {{\rm cum}}

\newcommand{\astr}      {^{\rm *}}

\begin{document}
\title{Signal Separation Using a Mathematical Model of Physiological Signals for the Measurement of Heart Pulse Wave Propagation With Array Radar}

\author{Takuya~Sakamoto
\thanks{This study was supported in part by JSPS KAKENHI 19H02155, JST PRESTO JPMJPR1873, and JST COI JPMJCE1307.}
\thanks{T.~Sakamoto is with the Department of Electrical Engineering, Graduate School of Engineering, Kyoto University, Kyoto, Kyoto 615-8510, Japan. sakamoto.takuya.8n@kyoto-u.ac.jp}
\thanks{T.~Sakamoto is also with Japan Science and Technology Agency, PRESTO, Kawaguchi, Saitama 332-0012, Japan.}}

\markboth{}{}

\maketitle
\begin{abstract}
    The arterial pulse wave, which propagates along the artery, is an important indicator of various cardiovascular diseases. By measuring the displacement at multiple parts of the human body, pulse wave velocity can be estimated from the pulse transit time. This paper proposes a technique for signal separation using an antenna array, so that pulse wave propagation can be measured in a non-contact manner. The body displacements due to the pulse wave at different body parts are highly correlated, and cannot be accurately separated using techniques that assume independent or uncorrelated signals. The proposed method formulates the signal separation as an optimization problem, based on a mathematical model of the arterial pulse wave. The objective function in the optimization comprises four terms that are derived based on a small-displacement approximation, unimodal impulse response approximation, and a causality condition. The optimization process was implemented using a genetic algorithm. The effectiveness of the proposed method is demonstrated through numerical simulations and experiments.
\end{abstract}


\IEEEpeerreviewmaketitle

\section{Introduction}
\IEEEPARstart{A}{ccording} to reports from the National Center for Health Statistics of the Centers for Disease Control and Prevention in the United States, the number of adults diagnosed with heart disease was 30.3 million in 2018 \cite{cdc}. That accounts for 12.1\% of all adults, resulting in more than 600 000 deaths/year, making heart disease the leading cause of death in the United States. Constant monitoring of health status is important for the prevention and treatment of heart disease. Some signs of heart disease can be detected from pulse wave propagation along the arteries caused by the pulsation of the heart. In particular, pulse wave velocity (PWV), which is the velocity of the pulse wave, is related to blood pressure and vascular stiffness, and is important for early detection of signs of heart disease such as hypertension, arteriosclerosis, and myocardial infarction. Conventionally, a cuff-type contact sensor has been used to measure pulse waves. For example, PWV can be calculated by wearing multiple cuffs on the four limbs, measuring the pulse transit time (PTT) between the upper arm and ankle, and dividing the distance between these parts by the PTT. However, in contact-type measurements, it is necessary to attach multiple sensors to multiple body parts at the same time. That makes the measurement inconvenient and time-consuming, resulting in discomfort and a sense of restraint, which makes this contact measurement unsuitable for long-term, continuous monitoring.

Instead of such contacting sensors, if non-contacting radar-based measurement is introduced, PWV can be continuously measured for a long time. We review existing radar-based measurement of pulse waves. Buxi et al.~\cite{Buxi2015}--\cite{Buxi2018} attached a 1-GHz continuous wave (CW) radar antenna to the neck/chest and also attached an impedance cardiography sensor on the waist/shoulders to measure the PTT from the time difference between the signals. Ebrahim et al.~\cite{Ebrahim2019,Ebrahim2019SR} attached a 900-MHz CW radar antenna to the chest, and also attached a photoelectric plethysmography (PPG) sensor to the ear; they measured the PTT from the rise time difference between the radar and PPG signals. Kuwahara et al.~\cite{Kuwahara2019} measured PTT and PWV from the peak time difference between the 2.4-GHz CW radar and a piezoelectric pulse wave sensor attached to the finger. In these studies, radar systems and contact sensors were used together, to measure pulse wave propagation.

Next, we discuss the measurement of pulse waves using only radar, without other sensors. Lauteslager et al.~\cite{Lauteslager2019} used ultra-wideband (UWB) radar with a center frequency of 3.8 GHz and a bandwidth of 1.0 GHz; they placed a single antenna on six body parts sequentially, assuming stationarity of pulse wave propagation during the measurement. Tao et al.~\cite{Tao2007} used two wristwatch-type impulse radio UWB radars with a bandwidth of 4 GHz. They attached them to the arm and leg and calculated the PTT and PWV from the delay time of the radar signals. Tang et al.~\cite{Tang2018} used a wrist-worn device with a CW radar and a self-injection-locked radar. The user is supposed to keep the device in front of their chest to measure the pulse wave at the chest and wrist simultaneously. In that study, the displacement of the chest was measured using non-contacting radar, but the displacement of the wrist was measured using contact-type radar. None of the above three studies reported non-contact measurements of pulse waves because they attached radar antennas to the body surface.

Lu et al.~\cite{Lu2010b} measured the chest and calf simultaneously using two C-band radars and calculated the PTT. Vasireddy et al.~\cite{Vasireddy2019} installed two 24-GHz radar systems 15-cm away from the chest and legs, and calculated the PTT. Although they installed radar antennas without touching the body, realizing a non-contact measurement, the installation of antennas near multiple body parts means that the positions of the antennas must be adjusted depending on the position and posture of the person. By contrast, Michler et al.~\cite{Michler2019} used phased-array radar at a distance of 10 cm from the abdomen, to perform simultaneous measurements at two abdominal points 7.3 cm apart, using a single radar system. Because the phased-array radar transmits signals in a specific direction, set in advance, the position and posture of the person being evaluated must be restricted, which is also unsuitable for long-term measurements.

In this research, non-contacting measurement of pulse wave propagation is performed using a single radar system with array antennas, without imposing restrictions on the position and orientation of the person. Non-contact and unrestrained pulse wave measurement is realized by using array signal processing, to form appropriate beam patterns according to the position and posture of the person being tested. The main problem is that the body displacements due to the pulse wave in multiple parts of the body are highly correlated, although there is a time difference corresponding to the PTT. For this reason, techniques for non-correlated signals are not suitable. In the present research, we propose a method to separate the radar echoes from multiple body parts using a mathematical model of physiological signals based on prior knowledge. In particular, the skin displacement signals due to the pulse wave are modeled to formulate the measurement of PTT as an optimization problem. The proposed method is applied to simulated and experimental data to demonstrate its performance.

{
  In this paper, we use the following notations: Lowercase $a$ for scalars, uppercase $A$ for matrices, and boldface lowercase $\vct{a}$ for vectors. The complex conjugate is denoted by a superscript asterisk $a\astr$. $\Re$ and $\Im$ denote the real and imaginary parts of a complex number, respectively. $\angle a$ denotes the phase of a complex value $a$. $\mathbb{C}^m$ denotes an $m$-dimensional complex coordinate space. The matrix transpose is denoted by a superscript T, as in $A^\TT$; the matrix conjugate transpose is denoted by a superscript H, as in $A^\HT$. The inverse of matrix $A$ is denoted as $A^{-1}$. Symbol $\circ$ denotes the element-wise multiplication operator. $\mathrm{diag}\{\cdots\}$ denotes a diagonal matrix whose entries are all zero except for its main diagonal, where the diagonal entries starting in the upper left corner are given in the argument.}

{
  The definitions of `uncorrelated' and `independent' are as follows. Let $f(t)$ and $g(t)$ be stationary ergodic signals. If $f(t)$ and $g(t)$ are uncorrelated, $\int_{-\infty}^{\infty}f\astr(t)g(t)\dd t=0$ holds. Otherwise, the signals are correlated. If $f(t)$ and $g(t)$ are independent, $p_{f}(f)p_{g}(g)=p_{fg}(f,g)$ holds, where $p_{f}(f)$ and $p_{g}(g)$ are the probability distribution functions of $f$ and $g$, and $p_{fg}(f,g)$ is the joint probability distribution function of $f$ and $g$. Otherwise, the signals are dependent.}

\section{Measurement of Pulse Wave Propagation}
The typical PWV in healthy young people is 3 to 5 m/s, and if it exceeds 10 m/s, treatment is required. In a simplified model, PWV is represented as $v_\mathrm{PWV}=\sqrt{\beta P/2 \rho}$, where $\beta$ is a stiffness parameter, $P$ is arterial pressure, and $\rho$ is the blood density. Because this formula includes $\beta$ and $P$, which are related to arteriosclerosis and hypertension, PWV is an important index that reflects various signs of cardiovascular disease. Widely used measures include carotid-femoral PWV, which measures the carotid and femoral arteries, and brachial-ankle PWV (baPWV), which measures the brachial and ankle arteries. Because the baPWV is measured by wearing cuffs on the arms and legs, restraining the subject during measurement is unavoidable. To perform non-contact measurement using a radar system instead of the conventional contact type measurement, the displacement of the skin surface caused by pulse wave propagation is measured. If the skin displacement at two body parts on the pulse wave propagation path can be measured at the same time, the PTT can be measured from the time difference between the pulse wave signals. The PWV is obtained simply by dividing the distance between the two body parts by the PTT. As described in the previous section, in the past, the antenna position had to be adjusted according to the position and orientation of the person being tested, so the method was non-contacting but not unconstrained. In this paper, a non-contacting and non-constraint pulse wave propagation measurement is realized, by combining an array radar system and array signal processing.

\section{Pulse Wave Measurement Using Radar}
\subsection{Signal Model in Radar Measurement of the Pulse Wave}
A radar system with an antenna array is used to measure physiological signals. In particular, we use a multiple-input multiple-output (MIMO) array comprising $M_1$ and $M_2$ elements for transmitting and receiving, respectively. {Following \cite{Wang_added}--\cite{Wan_added}, we assume that the distance from the array to the targets is much greater than the array aperture; array elements are closely spaced so that the direction of arrival of each echo can be approximated as the same for all elements. In addition, we also assume that mutual coupling between array elements is negligible. Under these conditions, the system is modeled using a virtual array of $M=M_1M_2$ elements.} In particular, we assume that the virtual array is a uniform linear array with a spacing of $\lambda/2$, where $\lambda$ is the wavelength.

Let us assume that the number of body parts (hereafter called "targets") contributing to the reflection is $N$, and that $N\leq M$ is satisfied. The displacement of the $j$-th target is $d_j(t)$ as a function of time $t$. Strictly speaking, $d_j(t)$ is not the actual displacement, but the line-of-sight component of the displacement. The skin displacement vector is denoted by $\vct{d}(t)=[d_1(t),d_2(t),\cdots,d_N]^\TT$. {We assume that the transmitted signal is approximated as a narrow-band signal and that the time delay corresponding to the propagation path length can be expressed as a phase shift.} The echoes are phase-modulated by the displacement as $s_j(t)\propto\ee^{\jj 2kd_j(t)}$, where $k=2\pi/\lambda$ is the wave number. The echo vector is denoted by $\vct{s}(t)=[s_1(t),s_2(t),\cdots,s_N(t)]^\TT$. Let the propagation channel between the $i$-th element and the $j$-th target be $a_{i,j}$, and the propagation channel {matrix} be an $M\times N$ matrix $A=(a_{i,j})$. We assume a stationary propagation path; $A$ does not change during the measurement. The signal $x_i(t)$ is received at the $i$-th element, which forms a signal vector $\vct{x}(t)=[x_1(t),x_2(t),\cdots,x_M(t)]^\TT$. Then $\vct{x}(t)$ is expressed as
\begin{equation}
  \vct{x}(t)=A\vct{s}(t)+\vct{n}(t),
\end{equation}
where $\vct{n}(t)$ is an additive noise component. The propagation channel matrix $A$ is also called a 'mixing matrix.' {For example, when assuming free-space propagation and targets located in the far-field, propagation channel $a_{i,j}$ can be approximated as $a_{i,j}\simeq \xi(\theta_j) \ee^{\jj 2k l_j}\ee^{\jj k u_i \sin\theta_j}/l_j^2$, where $\xi(\theta_j)$ is a complex coefficient that depends on the direction of arrival $\theta_j$, and $l_j$ is the distance between the $j$-th target and the center of the array. In addition, $u_i$ is the $x$ coordinate value of the $i$-th element, where the array baseline is on the $x$-axis.}

The purpose of this study is to find an 'unmixing matrix' $W$, with which we can obtain an estimate of the echo $\hat{\vct{s}}(t)$ as
\begin{equation}
  \hat{\vct{s}}(t) = W\vct{x}(t)
\end{equation}
and an estimate of the displacement $\hat{\vct{d}}(t)$ as
\begin{equation}
  \hat{\vct{d}}(t) = \frac{1}{2k}\angle\hat{\vct{s}}(t).
  \label{eq3}
\end{equation}

{For example, in a noiseless case with $A$ being nonsingular and square ($M=N$), we can set $W=A^{-1}$ so that the signal $\vct{s}(t)=A^{-1}\vct{x}(t)$ can be completely restored, including the order and amplitude of the signal.} It is not necessary to obtain $\vct{s}(t)$ itself; ambiguity of permutation and constant multiplication is allowed. In other words, if $\alpha_i \ee^{\jj kd_i(t)}$ is obtained with an arbitrary complex constant $\alpha_i$, important parameters such as PTT and PWV are estimated. For simplicity, the norm of each echo $s_j(t)$ is assumed to be 1, i.e., $|s_j(t)|=1$. In addition, displacement $d_i(t)$ is defined so that $\angle s_j(t)=kd_i(t)$, which means that the phase is zero when there is no displacement. If these conditions are not satisfied, the complex coefficient is incorporated into the mixing matrix. For example, if the echo vector $\vct{s}'$ is not normalized, the mixing matrix $A'$ can be replaced by $A$ as
\begin{equation}
    \begin{array}{ccl}
  \vct{x}(t)&=&A'\vct{s}'(t)+\vct{n}(t)\\
  &=&A' \mathrm{diag}\{|s_1'(t)|,\cdots,|s_N'(t)|\} \vct{s}(t)+\vct{n}(t)\\
  &=&A\vct{s}(t)+\vct{n}(t),
    \end{array}
\end{equation}
where $A = A' \mathrm{diag}\{|s_1'(t)|,\cdots,|s_N'(t)|\}$ and $s_j(t)=s_j'(t)/|s_j'(t)|$. Then, the echo can be expressed using only its phase as $s_j(t)=\ee^{\jj 2kd_j(t)}$.

{Note that the received signal includes not only echoes from the target person, but also clutter from surrounding objects (e.g., the floor and bed). Because the objects other than the target person are assumed to be stationary, the Doppler shift of the clutter is zero. Exploiting this characteristic, the zero-Doppler (DC) components of the received signal are removed before processing.}

\subsection{Signal Separation Using Adaptive Beamforming}
The minimum variance distortionless response (MVDR) \cite{Capon}--\cite{JLi2003}, which is also known as the directionally constrained maximization of power method \cite{Takao}, is an adaptive beamforming technique that minimizes the power of the weighted sum of the signals $y(t)=\vct{w}^\HT\vct{x}(t)$ while maintaining a constant response in the desired direction, e.g., $\vct{w}^\HT\vct{a}=1$ and $\vct{a}$ is a mode vector for the desired direction. The MVDR method can suppress interfering signals if the source signals are uncorrelated, i.e., $\EE[s_i(t)s_j(t)]=0$ for $i\neq j$. If the estimate of a mode vector $\hat{\vct{a}}_j=[\hat{a}_{1,j},\hat{a}_{2,j},\cdots,\hat{a}_{M,j}]^\TT$ for the $j$-th target is given, the signal from the $j$-th target is estimated as $\hat{s}_j(t)=\vct{w}_j^\HT \vct{x}(t)$ using a MVDR weight 
\begin{equation}
  \vct{w}_j=\left(\hat{\vct{a}}_j^\HT R^{-1}\hat{\vct{a}}_j\right)^{-1}{R^{-1}\hat{\vct{a}}_j},
\end{equation}
where $R=\EE[\vct{x}\vct{x}^\HT]$ is a correlation matrix. The unmixing matrix $W$ is obtained as $W=[\vct{w}_1,\vct{w}_2,\cdots,\vct{w}_N]^\HT$.

As stated above, the MVDR method is simple to implement and is effective for suppressing interference if the source signals are uncorrelated. Pulse waves measured at multiple body parts are, however, highly correlated, which does not satisfy the conditions for the MVDR method. For correlated signals, the spatial-averaging technique \cite{SA} is often used with the MVDR method, to un-correlate the signals by reducing the size of the array. However, using this technique, the effective array size is diminished and degrees of freedom lowered. In addition, the technique cannot be applied to arbitrary arrays. For these reasons, we do not discuss the spatial-averaging technique in the following sections.

\subsection{Signal Separation Using Independent Component Analysis}
Independent component analysis (ICA) is a technique of blind signal separation, which decomposes a multivariate signal into multiple non-Gaussian components (except for one component) that are statistically independent of each other.
In particular, JADE (Joint Approximation Diagonalization of Eigenmatrices) \cite{Cardoso} is a type of ICA; JADE uses the fourth-moment signals to decompose signals into independent components. Let us assume that the input signals have been whitened in preprocessing; their mean is zero ($\EE\{x_i\}=0$ $(1\leq i\leq M)$) and they are uncorrelated $\EE\{x_ix_j\}=0$ $(1\leq i\neq j\leq M)$. We also assume that the probability distribution function of each component $s_i$ $(1\leq i\leq N)$ is symmetric; its odd moments are zero. Because the components are statistically independent, the fourth cross-cumulant
\begin{equation}
  \begin{array}{c}
  \cum(s_i,s_j,s_k,s_l)=\EE\{s_is_js_ks_l\}-\EE\{s_is_j\}\{s_ks_l\}\\
  -\EE\{s_is_k\}\{s_js_l\}-\EE\{s_is_l\}\{s_js_k\}
  \end{array}
\end{equation}
is obtained as 
\begin{equation}
  \cum(s_i,s_j,s_k,s_l)=\left\{
  \begin{array}{c}
    c_i\,(i=j=k=l)\\
    0\,(\mathrm{otherwise})
  \end{array}
  \right.
\end{equation}

Using a matrix $M_0=(m_{ij})$, the cross-cumulant is reduced to a matrix
\begin{equation}
F_{i,j}(M_0)=\sum_{k,l=1}^{M}m_{k,l}\cum(x_i,x_j,x_k,x_l).
\end{equation}
If the unmixing matrix is unitary $W=(\vct{u}_1,\cdots,\vct{u}_N)$, we obtain
\begin{equation}
  F(M_0)=\sum_{k,l=1}^{M}(c_i \vct{u}_i^\TT M_0 \vct{u}_i)\vct{u}_i\vct{u}_i^\TT
\end{equation}
because $\cum(s_i,s_j,s_k,s_l)=0$ unless the indices are all equal.
By defining a diagonal matrix $D$ as $D(M)=\diag(c_i\vct{u}_1^\TT M_0\vct{u}_1,\cdots,c_M\vct{u}_M^\TT M_0\vct{u}_M,\cdots)$, we can express $F(M_0)$ as $F(M_0)=WD(M)W^\TT$. From this equation, it is found that the diagonalization of $F(M_0)$ gives the unmixing matrix $W$. By obtaining an eigenmatrix expansion of the tensor $\cum(x_i,x_j,x_k,x_l)$, we obtain eigenmatrices that are used to obtain multiple reduction matrices. We then find a unitary matrix $W$ that simultaneously diagonalizes the reduction matrices generated from the eigenmatrices using the Jacobi method. 

\section{Signal Separation Using a Mathematical Model of Physiological Signals}
In this section, we propose a mathematical model of a body displacement due to a pulse wave. Using the mathematical model, we formulate the signal separation process as an optimization problem. The objective function is proposed to estimate an unmixing matrix $W = [\vct{w}_1 \vct{w}_2 \cdots \vct{w}_N]^\TT$. For simplicity, each weight vector is assumed to be normalized as $|\vct{w}_i|=1$.

\subsection{Small-Displacement Approximation}
The displacement caused by pulse waves is, typically, at most 100 $\mu$m, although its value depends on individual differences and body parts. We assume that the displacement is sufficiently small, compared with the wavelength $\lambda=3.8$ mm corresponding to 79 GHz, which can be written as $2kd_j(t)=4\pi d_j(t)/\lambda \ll 2\pi$. Therefore, an echo can be approximated as
\begin{equation}
  \begin{array}{ccl}
    s_j(t)&=&\ee^{\jj 2kd_j(t)}\\
    &\simeq&1+\jj 2kd_j(t),    \label{eqflat}
  \end{array}
\end{equation}
where a displacement can be estimated without the $\angle$ operation. In addition, Eq.~(\ref{eqflat}) indicates that the I-Q plot of the echo on the complex plane can be approximated by a line segment, which can be expressed using the covariance matrix of the real and imaginary parts of the signal.

For the $i$-th echo, the covariance matrix $R_i$ of the real and imaginary parts of the complex-valued signal is written as
\begin{equation}
  R_i\!=\!\int_{-\infty}^{\infty}\left[
    \begin{array}{cc}
      \Re[\hat{{s}}_i(t)]^2 & \Re[\hat{{s}}_i(t)]\Im[\hat{{s}}_i(t)]\\
      \Im[\hat{{s}}_i(t)]\Re[\hat{{s}}_i(t)] & \Im[\hat{{s}}_i(t)]^2
    \end{array}
    \right]\dd t,
\end{equation}
where $\hat{{s}}_i(t) = \vct{w}^\mathrm{T}_i\vct{x}(t)$ and we should note that $R_i$ depends on $\vct{w}_i$. Let $\kappa(R_i)=|{\lambda_\mathrm{max}}/{\lambda_\mathrm{min}}|$ be the condition number of the matrix $R_i$, where $\lambda_\mathrm{max}$ and $\lambda_\mathrm{min}$ are the largest and smallest eigenvalues of $R_i$, respectively. To obtain echo estimates that can be approximated by Eq.~(\ref{eqflat}), we find ${\vct{w}_i\in\mathbb{C}^m}$ that maximizes $\kappa(R_i)$ for $i$.
Because $R_i$ is a $2\times 2$ matrix, its condition number can be written in a closed form as
\begin{equation}
  \kappa(R_i)=(\lambda_1(i)+\lambda_2(i))/(\lambda_1(i)-\lambda_2(i)),
\end{equation}
where
\begin{equation}
  \lambda_1(i) = \int_{-\infty}^{\infty} \left(\Re[\hat{{s}}_i]^2 + \Im[\hat{{s}}_i]^2\right)\dd t
\end{equation}
and
\begin{equation}
  \lambda_2(i)^2\!=\!\left|\int\!\Re[\hat{{s}}_i]^2\!-\!\Im[\hat{{s}}_i]^2\dd t\right|^2\!+\!4\left|\int \Re[\hat{{s}}_i]\Im[\hat{{s}}_i]\dd t\right|^2,
\end{equation}
where the integration interval $[-\infty, \infty]$ is omitted. We should note that $\lambda_1(i)$ and $\lambda_2(i)$ are both positive and $\lambda_1(i)>\lambda_2(i)$. Therefore, when the echo power $\lambda_1$ is constant, $\vct{w}_i$ that maximizes $\kappa(R_i)$ also maximizes $\lambda_2$. The same discussion is valid for all $i$ $(i=1,2,\cdots,n)$; the optimum unmixing matrix $W$ should maximize an objective function $F_1(W)$ defined as
\begin{equation}
  F_1(W)=\min_{1\leq i\leq n}\lambda_2(i)^2.
\end{equation}
By increasing $F_1(W)$, the estimated echo $\hat{s}_i(t)$ on the complex plane becomes flat and elongated, which is consistent with the small-displacement approximation in Eq.~(\ref{eqflat}). However, this model alone cannot estimate multiple echoes from different parts of a human body. In the next section, we extend the objective function so that the problem can be avoided.

\subsection{Simplified Pulse Wave Propagation Model}
In this section, we consider the relationship between the displacements at multiple body parts. We introduce a simplified model of pulse wave propagation; the displacements caused by the pulse wave are assumed to have similar waveforms with a time delay corresponding to the pulse transit time.

First, we assume that each displacement $d_i(t)$ $(i=1,2,\cdots,N)$ is approximated by a constant multiple of a time-shifted template waveform $d_0(t)$, i.e., $d_i(t)=\alpha_i d_0(t-\tau_i)$. We also assume that delay $\tau_i > 0$ holds for all $i$, and the delays are sorted in an ascending order as $\tau_1<\tau_2<\cdots <\tau_N$ without losing the generality of the model. With this assumption, body parts $1,2,\cdots,N$ are arranged in order from the heart side to the terminal side if the PWV is constant. The proposed model is formulated as
\begin{equation}
  \begin{array}{ccl}
  \vct{d}(t)&\simeq& \left[
    \begin{array}{c}
      \alpha_1 d_0(t-\tau_1)\\
      \alpha_2 d_0(t-\tau_2)\\
      \vdots\\
      \alpha_N d_0(t-\tau_N)
\end{array}
    \right]\\
  &=&d_0(t)*\left[
    \begin{array}{c}
      \alpha_1 \delta(t-\tau_1)\\
      \alpha_2 \delta(t-\tau_2)\\
      \vdots\\
      \alpha_N \delta(t-\tau_N)      
    \end{array}
    \right],
  \end{array}
\end{equation}
where $\delta()$ is the Dirac delta function and $*$ represents a convolution integral. The transfer function between $d_i(t)$ and $d_j(t)$ is obtained by deconvoluting the $j$-th component with the $i$-th component, and its impulse response $g_{i,j}(\tau)$ is calculated as follows:

\begin{equation}
  g_{i,j}(\tau) = \int_{-\infty}^{\infty}\frac{\int_{-\infty}^{\infty}d_j(t')\ee^{-\jj\omega t'}\dd t'}{\int_{-\infty}^{\infty}d_i(t')\ee^{-\jj\omega t'}\dd t'}\ee^{\jj \omega \tau}\dd \omega.
\end{equation}

If the above-mentioned conditions are satisfied, the impulse response $g_{i,j}(\tau)$ is approximated by
\begin{equation}
  g_{i,j}(\tau) \propto \delta(\tau+\tau_i-\tau_j).
\end{equation}
Strictly speaking, the waveforms of $d_i(t)$ and $d_j(t)$ do not match perfectly, and, as a result, $g_{i,j}(\tau)$ is not a delta function. To make $g_{i,j}(\tau)$ as similar to a delta function as possible, the fourth moment of $g_{i,j}(\tau)$ is also included in our objective function as
\begin{equation}
F_2(W)=\prod_{1\leq i<j \leq n}\frac{\int_{-\infty}^{\infty}\left|g_{i,j}(\tau)\right|^4\dd \tau}{\left(\int_{-\infty}^{\infty}\left|g_{i,j}(\tau)\right|^2\dd \tau\right)^2}.
\end{equation}

In addition to $F_2(W)$, we introduce another function $F_3(W)$ that takes a large value if $g_{i,j}(\tau)$ has a non-zero component only for $\tau>0$, which assumes a model expressed as $g_{i,j}(\tau)\propto \delta(\tau+\tau_i-\tau_j)$ and $\tau_j > \tau_i$.
\begin{equation}
F_3(W)=\prod_{1\leq i<j \leq n}\frac{\max_{\tau>0}\left|g_{i,j}(\tau)\right|^2}{\max_{\tau<0}\left|g_{i,j}(\tau)\right|^2}.
\end{equation}

\subsection{Orthogonality of Array Factors}
Finally, the beam patterns (array factors) for weight vectors $\vct{w}_1, \vct{w}_2, \cdots$ preferably are orthogonal because each weight vector is designed to receive a specific echo, whereas the other weights are designed to reject the echo. This can be formulated as
\begin{equation}
  F_4(W)= \prod_{1\leq i<j \leq n} \min\{(\vct{u}_i^\TT\vct{u}_j)^{-1}, \gamma \},
  \label{F4}
\end{equation}
where $\gamma$ is a constant, and $\vct{u}_i$ is a vector representing the beam pattern for the weight $\vct{w}_i$ expressed as 
\begin{equation}
  \vct{u}_i = (Q_\mathrm{DFT}\vct{w}_i)\circ (Q_\mathrm{DFT}\vct{w}_i)\astr
\end{equation}
where $Q_\mathrm{DFT}$ is a discrete Fourier transform matrix. In Eq.~(\ref{F4}), $\gamma$ was introduced to prevent $F_4$ from excessively increasing when the two vectors $\vct{u}_i$ and $\vct{u}_j$ are orthogonal.

To combine the four functions introduced above, we propose the following optimization problem for the measurement of the pulse wave at multiple body parts.
\begin{equation}
  \max_{W\in\mathbb{C}^{n\times m}}F(W),
\end{equation}
where
\begin{equation}
  F(W)=F_1(W)F_2(W)F_3(W)F_4(W).
\end{equation}

By maximizing the objective function $F(W)$, the optimal unmixing matrix $W$ is obtained. Then, the $W$ we obtain is used to estimate the echo $\hat{s}_i(t)$ and displacement $\hat{d}_i(t)$, and we estimate the PTT and PWV from $g_{i,j}(\tau)$.

\section{Performance Evaluation of Proposed Method in a Simulation}
\subsection{Simulation Model and Parameters}
\begin{figure}[tb]
    \begin{center}
      \epsfig{file=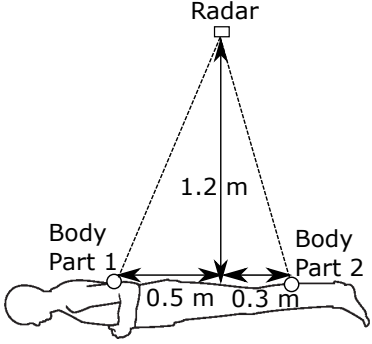,width=0.55\linewidth}
    \end{center}
  \caption{System model with the radar and human in a prone position. Body parts 1 and 2 are targets in our simulation.}
  \label{fig1}
\end{figure}

In this section, the performance of the conventional and proposed methods are evaluated through numerical simulations, which assumes a frequency of 79 GHz and an array of $M=12$ elements with element spacings of $\lambda/2$. The array is installed 1.25 m from the human body and the array baseline is aligned in the median plane of the body. For simplicity, $N=2$ is assumed; only two parts of the body contribute to the scattering of echoes. Body parts 1 and 2 are located at distances of 0.5 m and 0.3 m from the bottom of the antenna (see Fig.~\ref{fig1}), i.e., $x_1=-0.5$ m and $x_2=0.3$ m. From the antenna array, the nadir angles of parts 1 and 2 are $-22.6^\circ$ and $14.0^\circ$, respectively.

Here, the power values of echoes from parts 1 and 2 are assumed to be $P_1=0$ dB and $P_2=-3$ dB and the noise power is assumed to be $P_\mathrm{N}=-45$ dB; the signal-to-noise power ratio (S/N) for parts 1 and 2 are 45 dB and 42 dB. The S/N values do not reflect the actual measurement of pulse waves because we are interested only in the time-varying components of the echoes that contain physiological information. The displacement due to the pulse wave is $d_\mathrm{max} = \pm 50$ $\mu\mathrm{m}$, corresponding to a phase rotation of 19.0$^\circ$. In this simulation, the power of the time-varying component is lower than the time-invariant component by $20.43$ dB, resulting in effective S/N ratios of 24.6 dB and 21.6 dB, respectively. We set $\gamma = 10$ and the total measurement time $T$ was set to $T=20$ s.

\begin{figure}[tb]
    \begin{center}
      \epsfig{file=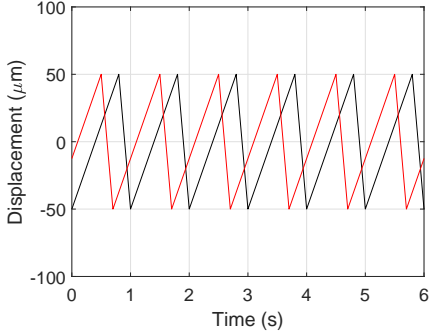,width=0.7\linewidth}
    \end{center}
  \caption{Simulated displacement waveforms $d_1(t)$ (red) and $d_2(t)$ (black) of body parts 1 and 2.}
  \label{fig2}
\end{figure}

\begin{figure}[tb]
  \hspace{-2mm}
  \begin{minipage}{0.49\hsize}
    \begin{center}
      \epsfig{file=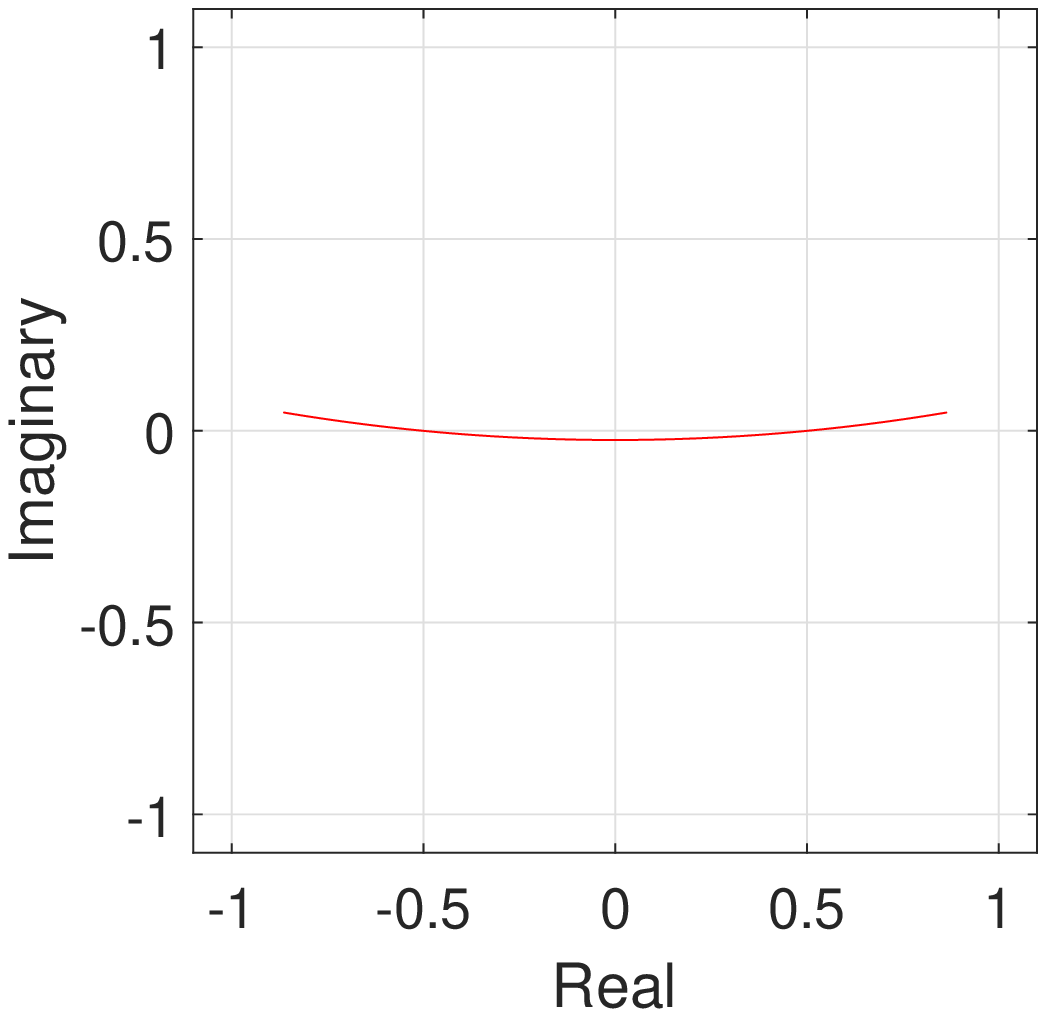,width=1.2\linewidth}
    \end{center}
  \end{minipage}
  \hspace{-2mm}
  \begin{minipage}{0.49\hsize}
    \begin{center}
      \epsfig{file=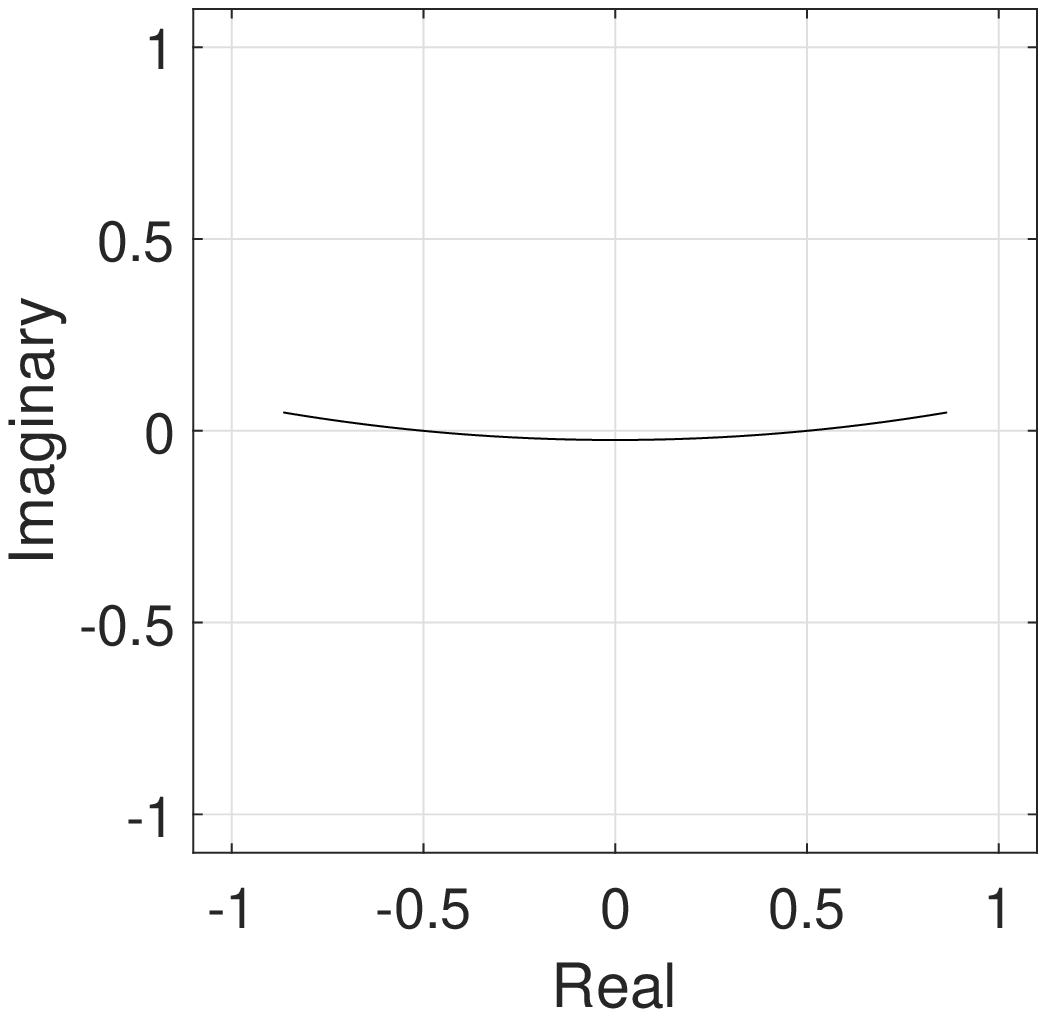,width=1.2\linewidth}        
    \end{center}
  \end{minipage}      
  \caption{I-Q plots of simulated echoes $s_1(t)$ (red) and $s_2(t)$ (black).}
  \label{fig2b}
\end{figure}

Fig.~\ref{fig2} shows the displacements $d_1(t)$ and $d_2(t)$ at body parts 1 and 2 assumed in the simulation, where the displacements are both triangular waves with a delay $\tau_2-\tau_1 = 300$ ms. Since body part 1 is closer to the heart than body part 2 and the distance between the parts is 0.8 m, it is about 2.7 m/s when converted to PWV. The left and right figures in Fig.~\ref{fig2b} show the I-Q plots of echoes from body parts 1 and 2 on the complex plane, both of which draw an arc as approximated in Eq.~(\ref{eqflat}).

\subsection{Performance Evaluation of Adaptive Array in Separating Pulse Wave Echoes in a Simulation}
In this section, we apply the Capon method \cite{Capon,JLi2003} and MVDR \cite{Takao} method to the simulated signal to evaluate their performance in separating radar echoes.

\begin{figure}[tb]
  \hspace{-2mm}
  \begin{minipage}{0.49\hsize}
    \begin{center}
      \epsfig{file=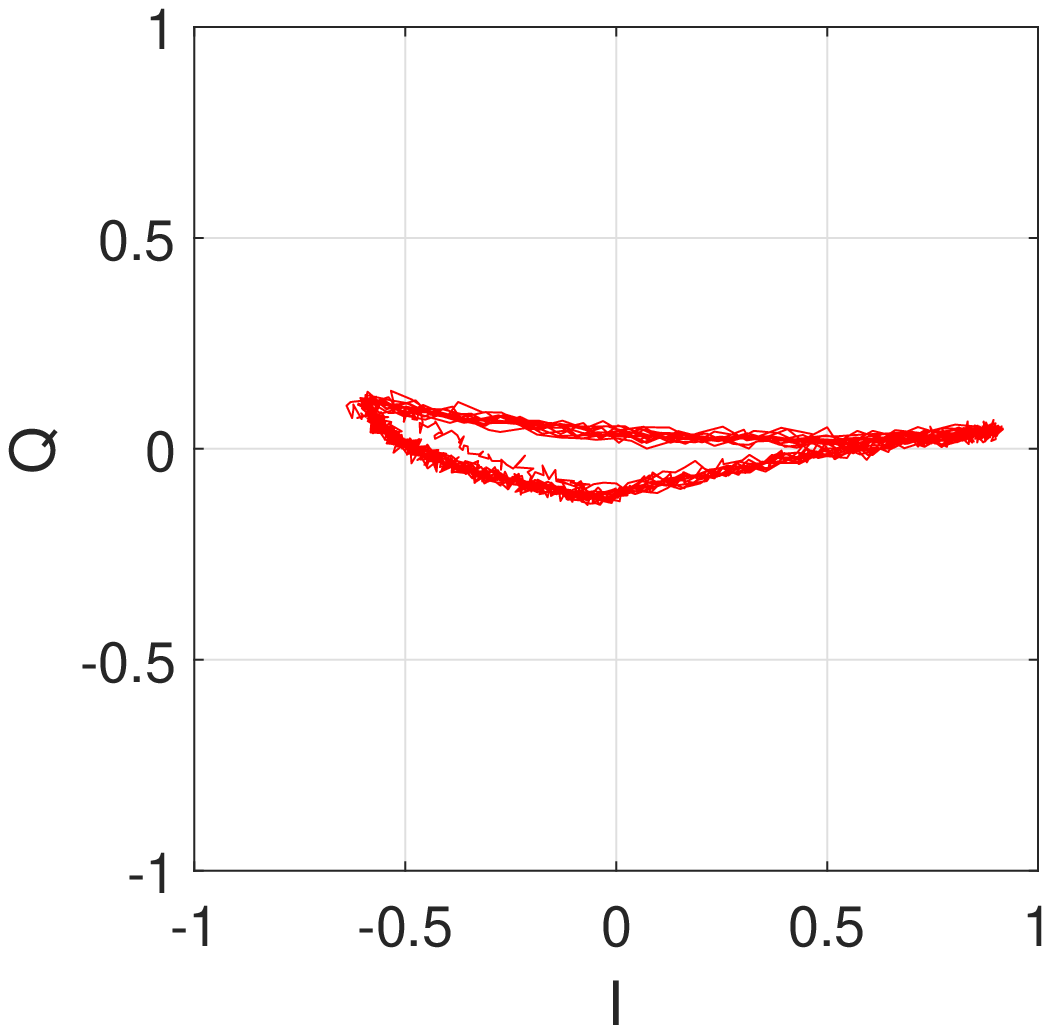,width=1.2\linewidth}
    \end{center}
  \end{minipage}
  \hspace{-2mm}
  \begin{minipage}{0.49\hsize}
    \begin{center}
      \epsfig{file=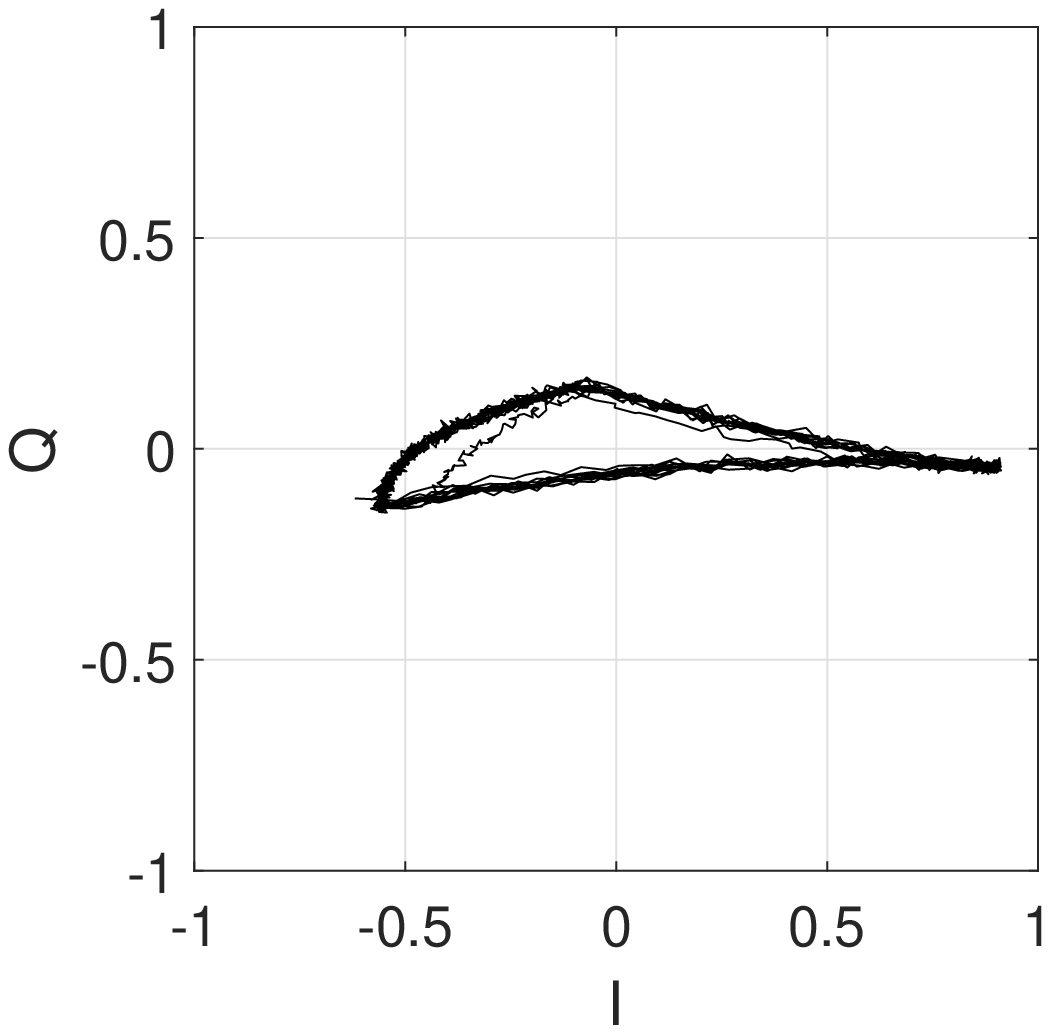,width=1.2\linewidth}        
    \end{center}
  \end{minipage}
    \caption{I-Q plots of complex-valued signals $\hat{s}_1(t)$ (left) and $\hat{s}_2(t)$ (right) estimated using the MVDR method.}
  \label{fig6}
\end{figure}

\begin{figure}[tb]
  \begin{center}
    \epsfig{file=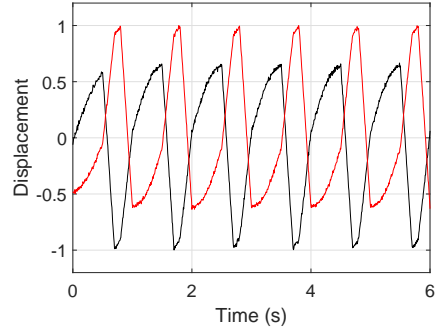,width=0.70\linewidth}
        \caption{Body displacement waveforms $\hat{d}_1(t)$ (red) and $\hat{d}_2(t)$ (black) estimated using the MVDR method.}
        \label{fig7}
    \end{center}
\end{figure}

\begin{figure}[tb]
  \begin{center}
    \epsfig{file=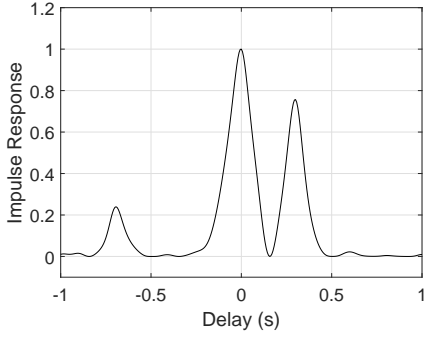,width=0.70\linewidth}
        \caption{Impulse response $g_{1,2}(\tau)$ estimated using the MVDR method.}
        \label{fig7b}
    \end{center}
\end{figure}

Fig.~\ref{fig6} shows the I-Q plots of the echoes estimated using the MVDR method. The I-Q plots look different from the ones in Fig.~\ref{fig2b}, which indicates that the signal separation has not been performed correctly. Fig.~\ref{fig7} shows the estimated displacement waveforms obtained from the signals separated by the MVDR method. The waveforms also seem different from the actual displacement in Fig.~\ref{fig2}. Because the displacement waveforms at multiple body parts are similar and they are correlated, the MVDR method does not work properly. The rms error $\epsilon_i$ in the displacement waveform $\hat{d}_i(t)$ is evaluated as
\begin{equation}
  \epsilon_i = \sqrt{\min_{\eta}\frac{1}{T}\int_0^{T} |d_i(t) - \eta \hat{d}_i(t)|^2 \dd t},
\end{equation}
where the coefficient $\eta$ is selected so that the error is minimized because we are interested only in the displacement waveform, not the scaling coefficient. For the estimation using the MVDR method, the rms errors $\epsilon_1=9.0$ $\mu$m and $\epsilon_2=10.1$ $\mu$m were obtained. Fig.~\ref{fig7b} shows impulse response $g_{1,2}(\tau)$ calculated from the displacements $\hat{d}_1(t)$ and $\hat{d}_2(t)$ obtained above. Although there are two peaks in $g_{1,2}(\tau)$, the peak close to the actual PTT is located at 305.5 ms. Because the actual PTT assumed in the simulation is 300 ms, the estimation error is $5.5$ ms and the relative error is 1.8\%.

\subsection{Performance Evaluation of Independent Component Analysis in Separating Pulse Wave Echoes in a Simulation}
\begin{figure}[tb]
  \hspace{-2mm}
  \begin{minipage}{0.49\hsize}
    \begin{center}
      \epsfig{file=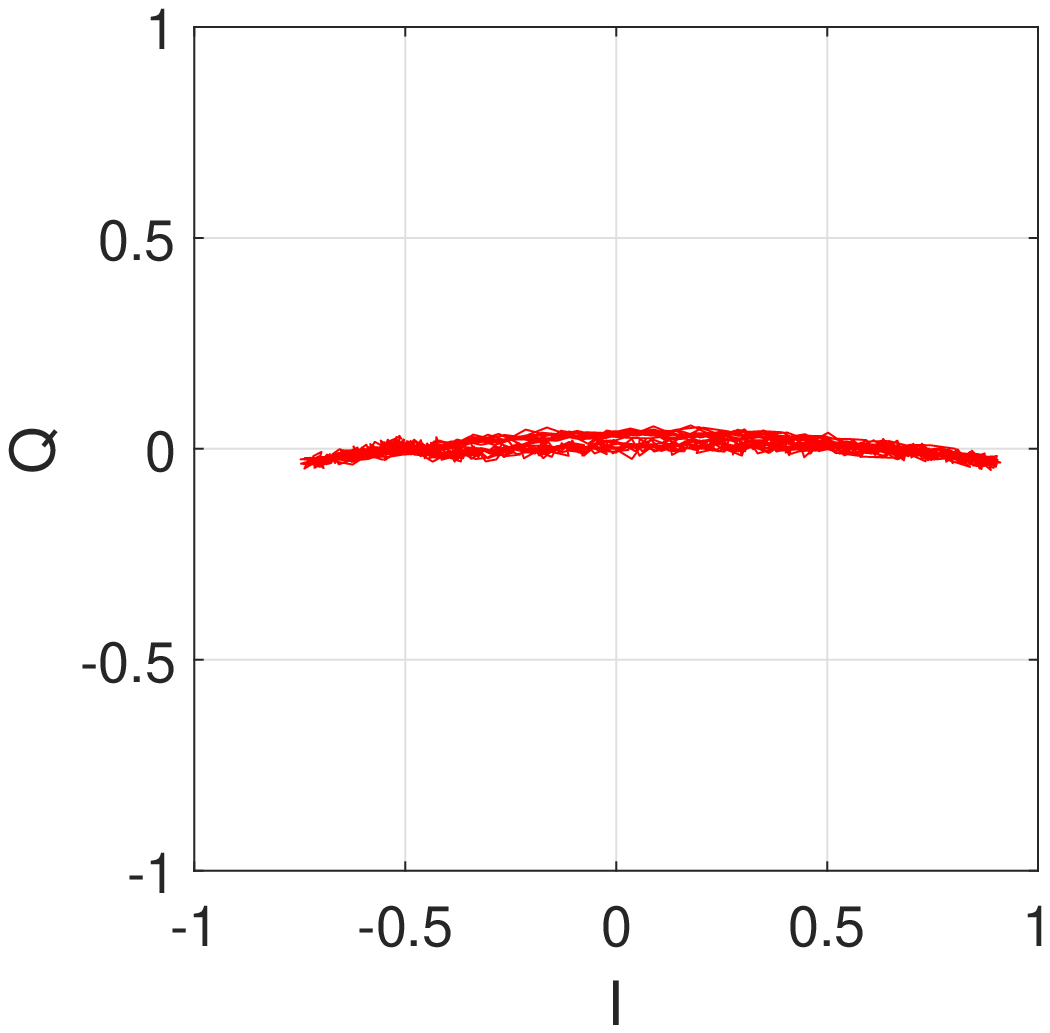,width=1.2\linewidth}
    \end{center}
  \end{minipage}
  \hspace{-2mm}
  \begin{minipage}{0.49\hsize}
    \begin{center}
      \epsfig{file=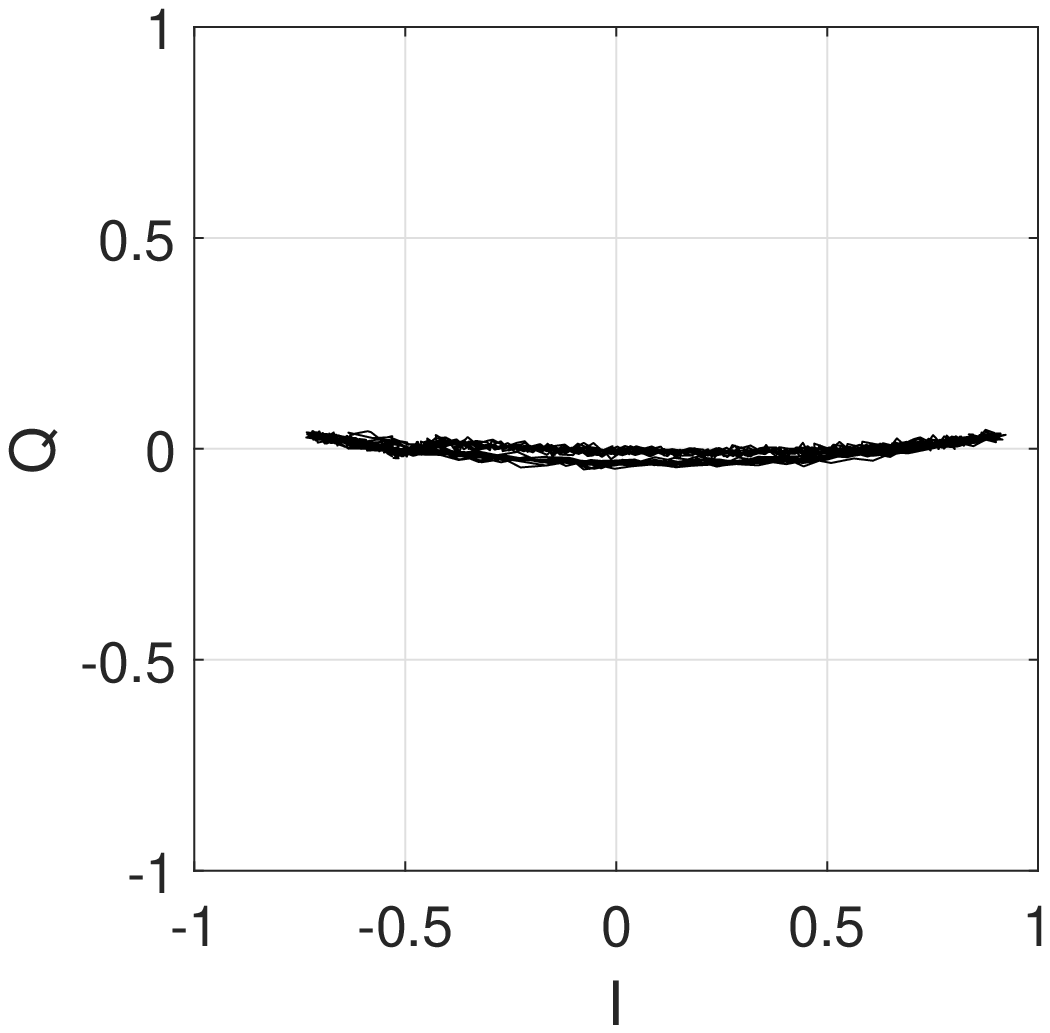,width=1.2\linewidth}        
    \end{center}
  \end{minipage}      
  \caption{I-Q plots of complex-valued signals $\hat{s}_1(t)$ (left) and $\hat{s}_2(t)$ (right) estimated using JADE-ICA.}
  \label{fig10}
\end{figure}

\begin{figure}[tb]
  \begin{center}
    \epsfig{file=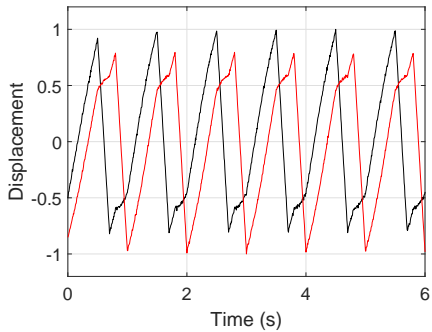,width=0.7\linewidth}
        \caption{Skin displacement waveforms $\hat{d}_1(t)$ (red) and $\hat{d}_2(t)$ (black) estimated using the JADE-ICA algorithm.}
        \label{fig11}
    \end{center}
\end{figure}

\begin{figure}[tb]
  \begin{center}
    \epsfig{file=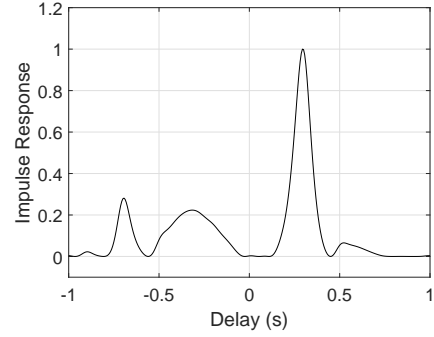,width=0.7\linewidth}
    \caption{Impulse response $g_{1,2}(\tau)$ estimated using the JADE-ICA algorithm.}
        \label{fig11b}
    \end{center}
\end{figure}

In this section, we apply JADE-ICA to the simulated signals. To investigate its performance under an ideal condition, the actual number of targets is given. Fig.~\ref{fig10} shows the I-Q plots of the estimated echoes. Compared with the estimates with the MVDR method (Fig.~\ref{fig6}), the trajectory seems closer to the actual one (Fig.~\ref{fig2b}). Fig.~\ref{fig11} shows the displacement waveforms $\hat{d}_1(t)$ and $\hat{d}_2(t)$ estimated by the JADE-ICA. Compared with the estimates using the MVDR method, the displacement waveforms in Fig.~\ref{fig11} look closer to the actual ones. Despite this, both of the waveforms are distorted, with rms errors of $\epsilon_1=5.6$ $\mu$m and $\epsilon_2=5.8$ $\mu$m. Fig.~\ref{fig11b} shows impulse response $g_{1,2}(\tau)$ calculated from the $\hat{d}_1(t)$ and $\hat{d}_2(t)$ obtained above. The peak of $g_{1,2}(\tau)$ is located at 294.7 ms, and the estimation error is $5.3$ ms and the relative error is 1.8\%.

\subsection{Performance Evaluation of the Proposed Method in Separating Pulse Wave Echoes in a Simulation}
Next, we investigate the performance of the proposed method in separating the simulated signals. The actual number of targets $N=2$ is given in the same way as in the evaluation of the MVDR method and JADE-ICA, and thus, the size of the unmixing matrix $W$ is $N\times M = 2\times 12$. The optimization problem
\begin{equation}
  W^{\astr}=\arg  \max_{W\in\mathbb{C}^{n\times m}}F(W)
  \label{eq:opt1}
\end{equation}
is solved to obtain an unmixing matrix $W^{\astr}$ to estimate an echo $\hat{\vct{s}}(t)=W^{\astr}\vct{x}(t)$. The estimated echo $\hat{\vct{s}}(t)$ is rotated so that its principal component is directed in the real axis, and its real part is output as the displacement estimate $\hat{\vct{d}}(t)$.

A genetic algorithm (GA) is used to maximize the objective function in Eq.~(\ref{eq:opt1}). Because the number of complex unknowns is $NM=24$ in $W$, a 24-dimensional vector of complex numbers is used as a gene in the GA. The elements of unmixing matrix $W$ are all initialized to be 1's. The number of individuals in each generation is 100, and the value of the objective function is used as the fitness. Roulette selection with a probability proportional to fitness is used for selection. In the crossover, the row vectors in the selected two-generation individual matrix are exchanged to generate the two next-generation individuals.

\begin{figure}[tb]
    \begin{center}
      \epsfig{file=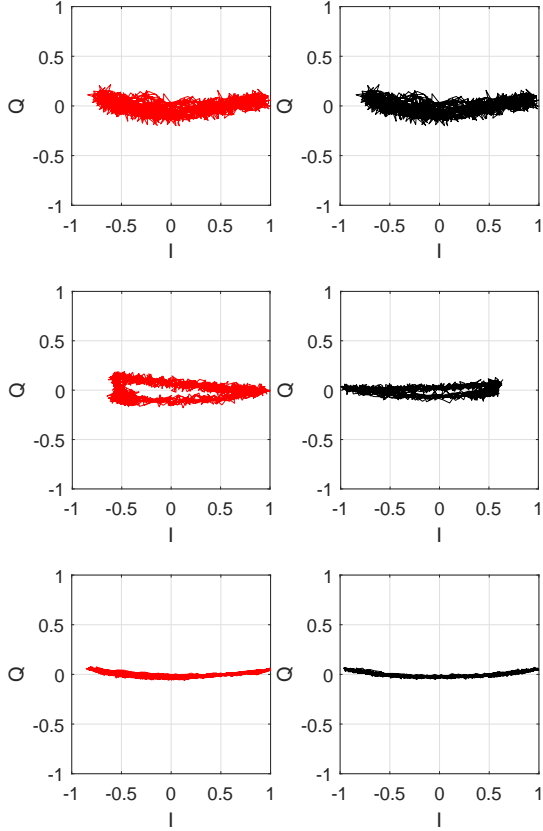,width=0.95\linewidth}
    \end{center}
    \caption{I-Q plots of complex-valued signals $\hat{{s}}_1(t)$ (red, left) and $\hat{{s}}_2(t)$ (black, right) estimated using the proposed method that maximizes $F(W)$. Shown are the estimations from the 1st (top), 3rd (middle), and 50th (bottom) generations of the genetic algorithm.}
  \label{fig14}
\end{figure}

\begin{figure}[tb]
\begin{center}
\epsfig{file=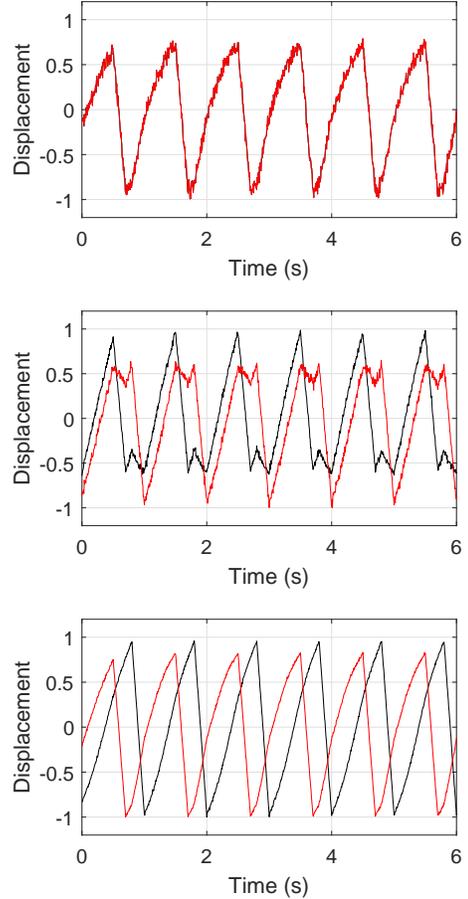,width=0.75\linewidth}
\end{center}
\caption{Body displacements $\hat{{d}}_1(t)$ (red line) and $\hat{{d}}_2(t)$ (black line) estimated using the proposed method that maximizes $F(W)$, for the 1st (top), 3rd (middle), and 50th (bottom) generations of the genetic algorithm.}
\label{fig15}
\end{figure}

\begin{figure}[tb]
\begin{center}
\epsfig{file=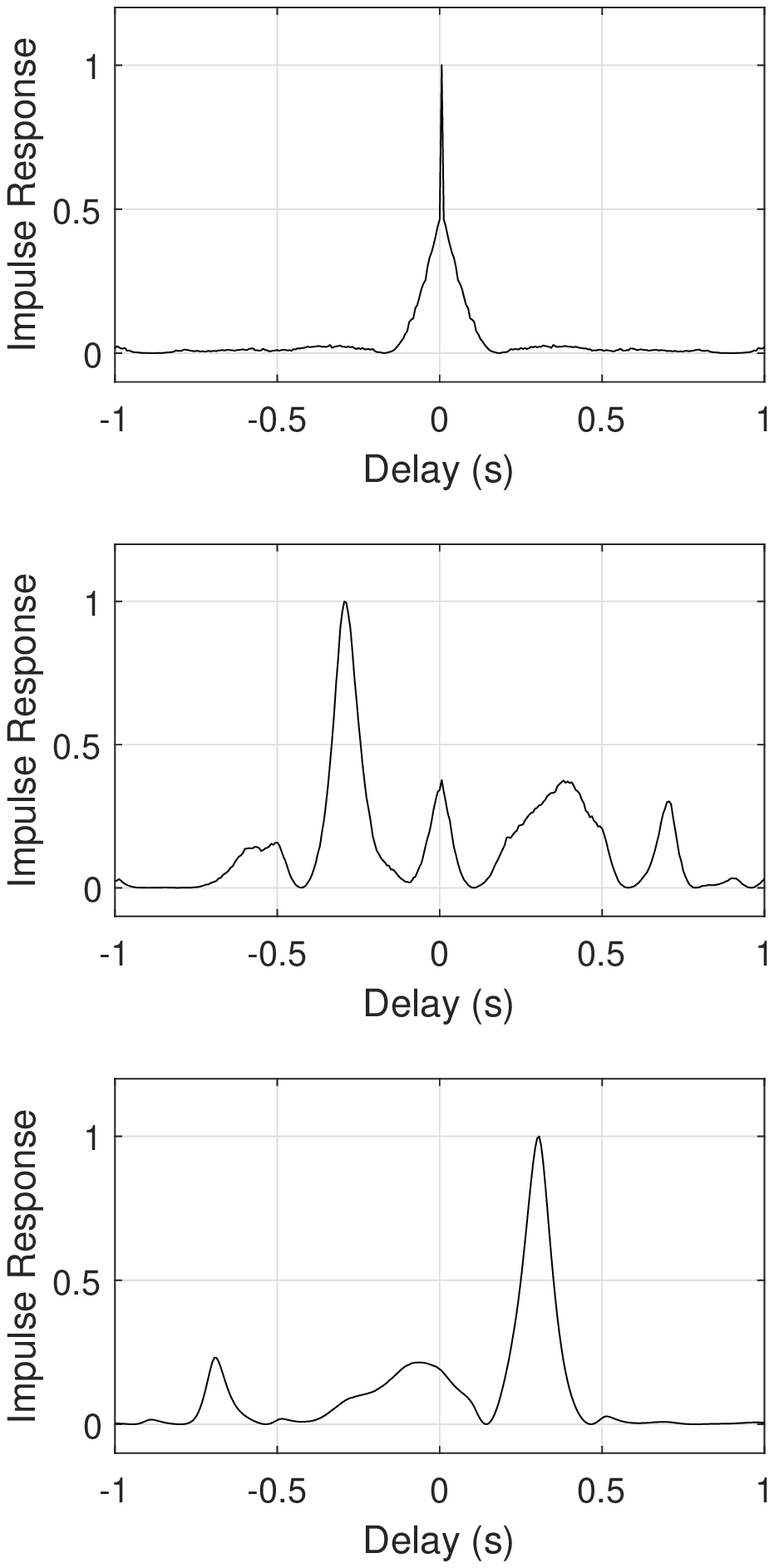,width=0.75\linewidth}
\end{center}
\caption{Impulse response $g_{1,2}(\tau)$ estimated using the proposed method that maximizes $F(W)$. Shown are the estimations from the 1st (top), 3rd (middle) and 50th (bottom) generations of the genetic algorithm.}
\label{fig16}
\end{figure}

Fig.~\ref{fig14} shows the I-Q plots of echoes $\hat{s}_1(t)$ (red line) and $\hat{s}_2(t)$ (black line) estimated from the unmixing matrix $W$ of the best individual in the 1st, 3rd, and 50th generations of the GA maximizing $F(W)$, where we see that the I-Q plots become flat by increasing $F_1(W)$ as intended. Fig.~\ref{fig15} shows body displacements $\hat{d}_1(t)$ (red line) and $\hat{d}_2(t)$ (black line) estimated in the 1st, 3rd, and 50th generations of the GA. The estimated body displacement in the 50th generation seems similar to the actual displacement shown in Fig.~\ref{fig2}, where the rms errors were $\epsilon_1=2.5$ $\mu$m and $\epsilon_2=4.7$ $\mu$m, indicating that the both errors are smaller than those for the MVDR method and JADE-ICA. Fig.~\ref{fig16} shows the impulse responses $g_{1,2}(\tau)$ obtained from these estimated displacements in the 1st, 3rd, and 50th generations of the GA-based optimization.

Because $\hat{d}_1(t)$ and $\hat{d}_2(t)$ are the same in the first generation, the impulse response $g_{1,2}(\tau)$ has its peak at $\tau=0$, which reduces the values of $F_3(W)$, lowering the value of the objective function {$F(W)$}. In the 50th generation, however, the impulse response $g_{1,2}(\tau)$ has its value at $\tau>0$, satisfying the causality condition, increasing the value of $F_3(W)$. Specifically, the peak of $g_{1,2}(\tau)$ in the 50th generation is located at $\tau=300.9$ ms, the error in estimating the PTT is $0.9$ ms and the relative error is 0.3\%, which indicates the higher accuracy of the proposed method compared with the conventional MVDR and ICA approaches.

\begin{figure}[tb]
\begin{center}
    \epsfig{file=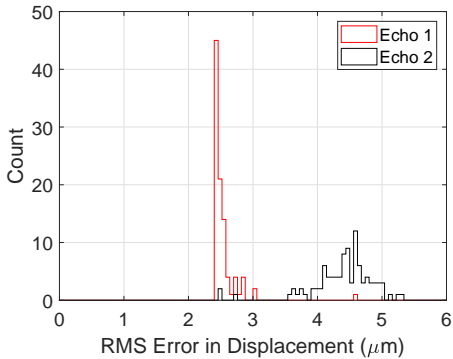,width=0.75\linewidth}
\end{center}
\caption{Histogram of the rms errors in displacements using the proposed method.}
\label{fig16b}
\end{figure}

\begin{figure}[tb]
\begin{center}
   \epsfig{file=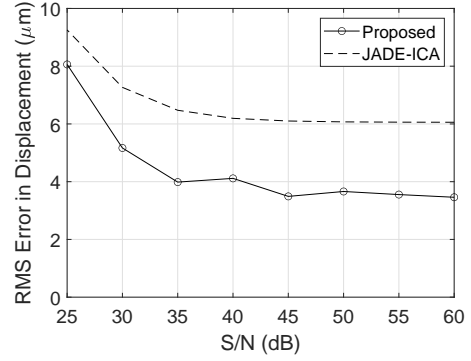,width=0.75\linewidth}
\end{center}
\caption{Average rms errors vs. S/N in estimating displacements using the ICA and the proposed method.}
\label{fig16c}
\end{figure}

We investigate the performance of the proposed method using multiple random seeds for generating pseudorandom numbers used in the GA. Fig.~\ref{fig16b} shows the histograms of the rms errors in estimating the displacements $d_1$ and $d_2$ using the proposed method, where 100 different random seeds were used. The mean$\pm$SD errors in estimating the displacements $d_1$ and $d_2$ were 2.9 $\mu$m$\pm$1.9 $\mu$m and 4.5 $\mu$m$\pm$1.1 $\mu$m, respectively, whereas using the MVDR and ICA, the average errors were 9.6 $\mu$m and 5.7 $\mu$m. Fig.~\ref{fig16c} shows the rms errors in estimating the displacements using the ICA and the proposed method. For each S/N, 100 random seeds were used, among which, {errors larger than 15 $\mu$m were excluded in calculating the rms errors. The large errors were caused by insufficient numbers of generations and population size in the GA. For $\mathrm{S/N}=25$ and 30 dB, 2 cases out of 100 resulted in large errors; for $\mathrm{S/N}=35$ and 50 dB, 1 case out of 100 resulted in large errors. The figure shows that the proposed method achieved a higher accuracy than did the ICA.}

Next, we investigate the performance of the 3 methods in another simulation with $x_1=-0.1$ m, $x_2=0.3$, and $P_1=P_2=0$ dB; the rest of the parameters were the same. For the proposed method, simulations were run with 100 random seeds. Among 100 random seeds, 3 seeds gave large PTT errors: 293.2, 293.0, and 294.0 ms, whereas the other random seeds gave small errors. Except for those 3 cases, the mean error in estimating the PTT was 0.57 ms$\pm$0.12 ms, whereas using the MVDR and ICA, the errors were 230.3 ms and 5.4 ms, respectively. As for the estimation of displacements, the errors were 19.7 $\mu$m, 6.1 $\mu$m, and 3.8 $\mu$m$\pm$0.57 $\mu$m using the MVDR, ICA, and the proposed method, respectively. These results demonstrate the effectiveness of the proposed method.

\subsection{Application of the Proposed Method to Experimental Data}
In this section, we apply the proposed method to experimental radar data from a participant. In the experiment, a 79-GHz millimeter-wave radar was used, as in the simulation in the previous section. The radar is a frequency-modulated CW system with a 99\%-occupied bandwidth of 3.5 GHz. The antenna is a MIMO array that uses three elements for transmission and four elements for reception; a total of seven elements are on the same straight line, and the transmission and reception arrays have element spacings of $2\lambda$ and $\lambda/2$, respectively. Therefore, the virtual array is a linear array with 12 elements with half-wavelength spacings, which is the same condition as the simulation in the previous section. 

\begin{figure}[tb]
    \begin{center}
      \epsfig{file=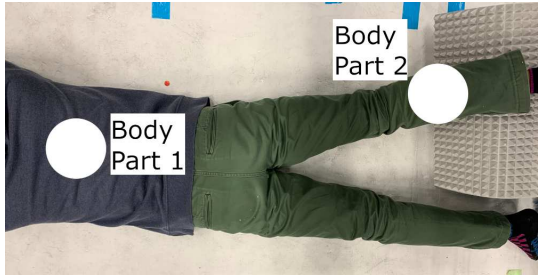,width=0.8\linewidth}
    \end{center}
  \caption{Photograph of a participant in the radar measurement.}
  \label{fig_pic}
\end{figure}

The origin of the $xyz$ Cartesian coordinates is the center of the array. The antenna array is in the $xy$ plane, in which the array baseline is on the $x$-axis. The downward vertical direction is the $z$ axis. The array is installed facing the floor at a height of 1.4 m from the top of the Styrofoam bed placed on the floor. The antenna elements have a linear polarization in the $y$ direction, and the element beam widths in the E plane ($yz$ plane) and H plane ($xz$ plane) are $8^\circ$ and $70^\circ$, respectively. When the beamforming method is used, the baseline length of the virtual array is $11\lambda/2$, so the beam width of the array factor in the H plane ($xz$ plane) is $9.3^\circ$. The participant lies on the bed in a prone position with his body aligned along the $x$ direction. The distance between the upper part of the participant and the array center was about 1.25 m. The footprint of the element beam was $152.7\times 17.5$ cm in the $x$ and $y$ directions, which covered the majority of the person's body, while suppressing echoes from the floor. Fig.~\ref{fig_pic} shows a photograph of the participant during the measurement. Targets 1 and 2 correspond to the back and calf, respectively. As a form of preprocessing, range gating was adopted; we removed signals from ranges other than that of the two body parts, to suppress clutter, as
\begin{equation}
\vct{x}(t)=\int_{r_1}^{r_2}\vct{x}'(r,t)\dd r,
\end{equation}
where $\vct{x}'(r,t)\dd r$ is a signal vector received for time $t$ and range $r$, and the two body parts are located within $r_1\leq r\leq r_2$, which was set manually.

\begin{figure}[tb]
    \begin{center}
      \epsfig{file=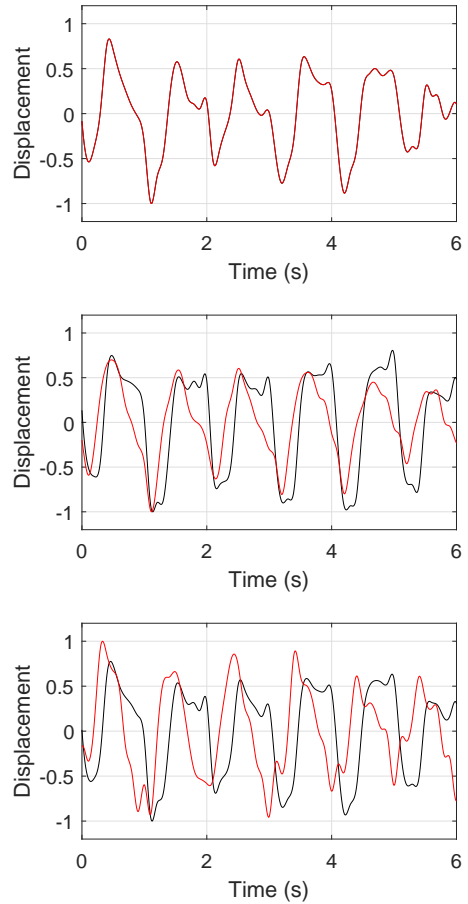,width=0.75\linewidth}
    \end{center}
    \caption{Body displacements $\hat{{d}}_1(t)$ (red) and $\hat{{d}}_2(t)$ (black) estimated using the proposed method applied to the experimental data. The estimations from the 1st (top), 3rd (middle) and 50th (bottom) generations of the genetic algorithm are shown.}
  \label{fig_ex2}
\end{figure}

\begin{figure}[tb]
    \begin{center}
      \epsfig{file=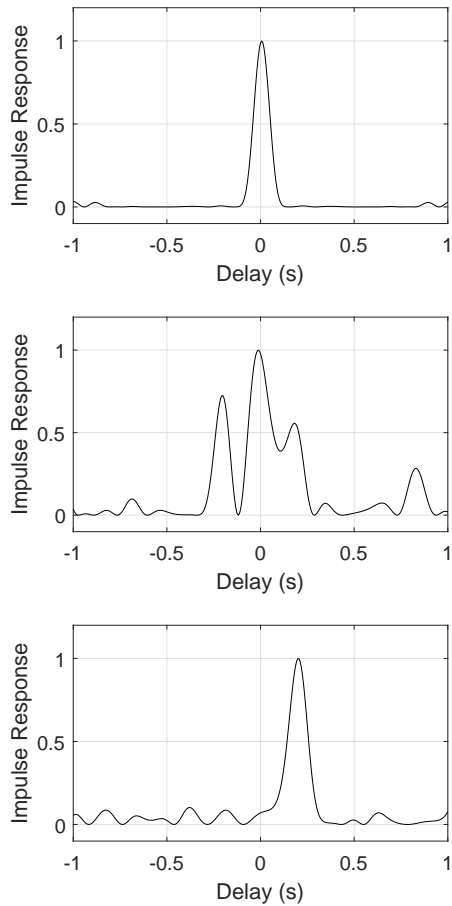,width=0.75\linewidth}
    \end{center}
    \caption{Impulse response $g_{1,2}(\tau)$ estimated using the proposed method applied to the experimental data. Shown are the estimations from the 1st (top), 3rd (middle), and 50th (bottom) generations of the genetic algorithm.}
  \label{fig_ex3}
\end{figure}

The proposed method was applied to the measured signal $\vct{x}(t)$, and used to investigate the method's performance in separating radar echoes with experimental data. Fig.~\ref{fig_ex2} shows the body displacements $\hat{d}_1(t)$ (red line) and $\hat{d}_2(t)$ (black line) estimated in the 1st, 3rd and 50th generations of the GA optimization, where we see that there is a delay between the displacement waveforms shown in red and black. The impulse responses $g_{1,2}(\tau)$ calculated from these waveforms are shown in Fig.~\ref{fig_ex3}. In the 1st generation, a peak appears at $\tau=0$, and in the 3rd generation, other peaks appear at $\tau\neq 0$. Finally, in the 50th generation, the main peak is shifted to $\tau > 0$. From the impulse response in the 50th generation, the propagation delay was estimated to be 173.1 ms. Because the distance between body parts 1 and 2 was about 75 cm, the PWV was calculated as 4.33 m/s. In the future, our next step is to verify the accuracy of the PWV estimated using the proposed method, by performing pulse wave measurements with other contact sensors and the radar simultaneously.

\section{Discussion}

\subsection{Posture of Target Person}
{In this study, we focused on the radar echoes from a target person in a prone position only. The posture of the person determines the body parts contributing radar echoes, which affects the number of sources, the directions of arrival, the necessary angular resolution and the skin displacement waveforms that are used in our proposed method. It will be important to study how the posture of the target person affects the performance in separating skin displacement signals and estimating PTT and PWV.}

\subsection{Methods for Separating Correlated Signals}
{Other than the proposed method, there are existing methods for separating correlated signals. One example is the use of compressed sensing (CS). In the on-grid CS method, possible values for each parameter to estimate are represented by a set of discrete grid points, where most of the grid values are zeros, and this property is exploited to separate signals. In recent works, in addition to the on-gird CS method, various off-grid and gridless CS methods have been proposed and demonstrated to achieve both high accuracy and robustness \cite{Zheng1_added,Zheng2_added}. The performance of these methods, as applied to our problem setting, will be investigated in our future work.}

\section{Conclusion}
In this study, we proposed a signal processing technique for non-contact measurement of the pulse wave propagating along the artery using an array radar system. The proposed technique allows us to measure the body displacement at multiple parts of the human body simultaneously. The echoes from multiple body parts are phase-modulated by a correlated displacement waveform, which lowers the performance of conventional methods such as the MVDR method and JADE-ICA. We proposed a mathematical model of body displacements caused by the pulse waves. The model is based on a small-displacement approximation and time-shift waveform approximation of the pulse waves. Based on this model, we formulated the signal separation as an optimization problem. Then, a GA was adopted to maximize the objective function and obtain an unmixing matrix, to estimate the echoes from multiple body parts. We also conducted an experiment with a participant using a 79-GHz, 12-channel MIMO array radar. Our study demonstrated that the accuracies in estimating the body displacements and PTT were both improved by the proposed method, compared with the conventional MVDR and JADE-ICA methods, which indicates that the proposed method is a promising approach to achieve an accurate non-contact PWV measurement using array radar.

\newpage
\begin{IEEEbiography}[{\includegraphics[width=1in,height=1.25in,clip,keepaspectratio]{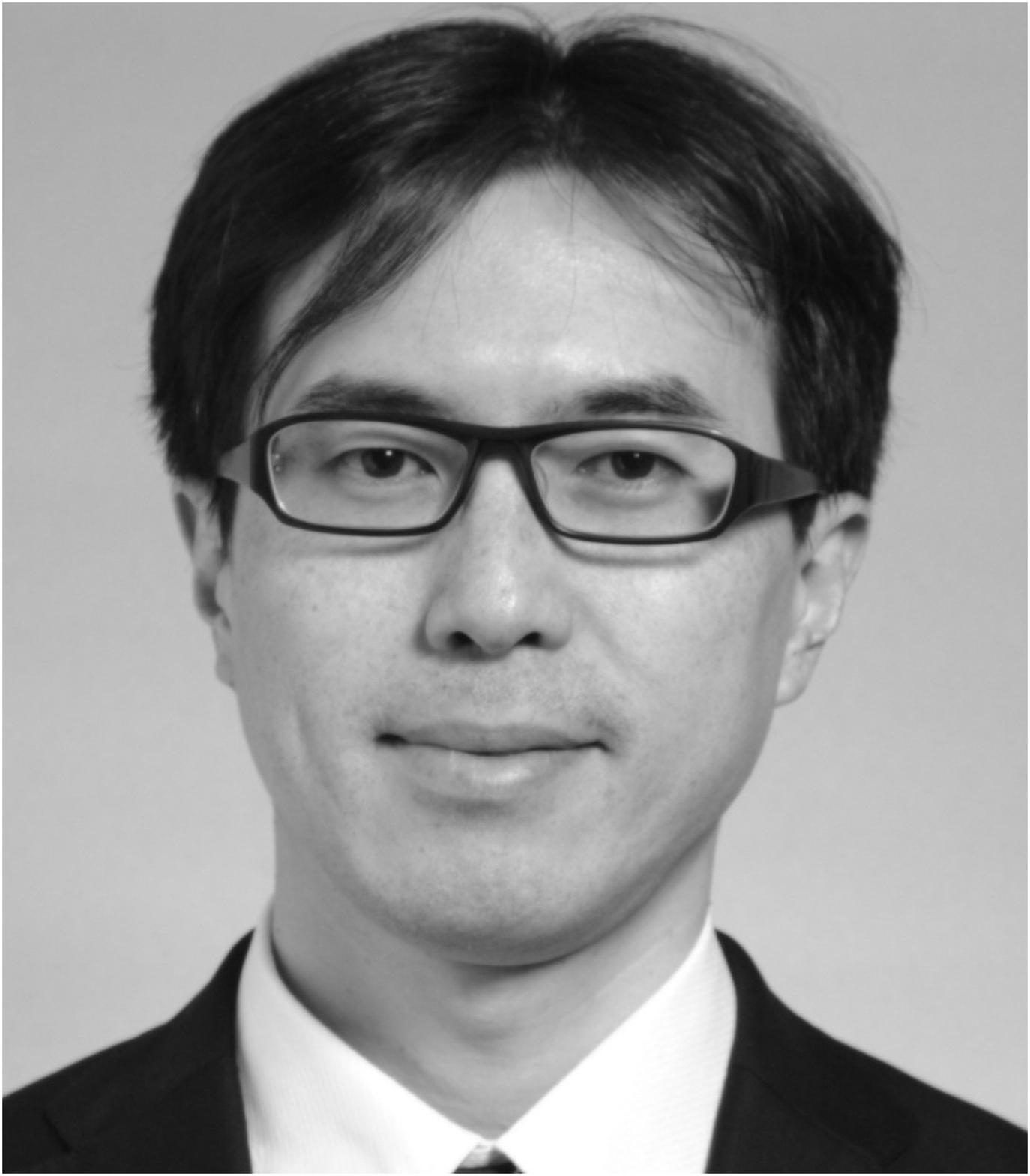}}]{Takuya Sakamoto}
(M'04-SM'17) received a B.E.~degree in electrical and electronic engineering from Kyoto University, Kyoto, Japan, in 2000 and M.I.~and Ph.D.~degrees in communications and computer engineering from the Graduate School of Informatics, Kyoto University, in 2002 and 2005, respectively. 

From 2006 to 2015, he was an Assistant Professor at the Graduate School of Informatics, Kyoto University. From 2011 to 2013, he was also a Visiting Researcher at Delft University of Technology, Delft, the Netherlands. From 2015 to 2018, he was an Associate Professor at the Graduate School of Engineering, University of Hyogo, Himeji, Japan. In 2017, he was also a Visiting Scholar at the University of Hawaii at Manoa, Honolulu, HI, USA. From 2018, he has been a PRESTO Researcher at the Japan Science and Technology Agency, Kawaguchi, Japan. Currently, he is an Associate Professor at the Graduate School of Engineering, Kyoto University. His current research interests are system theory, inverse problems, radar signal processing, radar imaging and wireless sensing of vital signs.

Dr. Sakamoto was a recipient of the Best Paper Award from the International Symposium on Antennas and Propagation (ISAP) in 2012, and the Masao Horiba Award in 2016. In 2017, he was invited as a semi-plenary speaker to the European Conference on Antennas and Propagation (EuCAP) in Paris, France. 
\end{IEEEbiography}

\end{document}